\documentclass[prl,twocolumn,floatfix,nobibnotes,footinbib,superscriptaddress,showpacs]{revtex4}
\usepackage{amsmath}
\usepackage{graphicx}
\usepackage{color}

\newcommand{\vc}[1]{\ensuremath{\mathbf{#1}}}

\begin{document}


\title{Efficient perturbation theory for quantum lattice models}

\author{H. Hafermann}
\affiliation{I. Institute for Theoretical Physics, University of Hamburg, 20355 Hamburg, Germany}
\author{G. Li}
\affiliation{Bethe Center for Theoretical Physics, University of Bonn, 53115 Bonn, Germany}
\author{A. N. Rubtsov}
\affiliation{Department of Physics, Moscow State University, 119992 Moscow, Russia}
\author{M. I. Katsnelson}
\affiliation{Institute for Molecules and Materials, Radboud
University of Nijmegen, 6525 ED Nijmegen, The Netherlands}
\author{A. I. Lichtenstein}
\affiliation{I. Institute for Theoretical Physics, University of Hamburg, 20355 Hamburg, Germany}
\author{H. Monien}
\affiliation{Bethe Center for Theoretical Physics, University of Bonn, 53115 Bonn, Germany}

\begin{abstract}
We present a novel approach to long-range correlations beyond dynamical mean-field theory through a ladder approximation to dual fermions. The new technique is applied to the two-dimensional Hubbard model. We demonstrate that the transformed perturbation series for the nonlocal dual fermions has superior convergence properties over standard diagrammatic techniques.
The critical N\'eel temperature of the mean-field solution is suppressed in the ladder approximation, in accordance with quantum Monte-Carlo (QMC) results. An illustration of how the approach captures and allows to distinguish short- and long-range correlations is given.
\end{abstract}
\maketitle

The physics of strongly interacting quantum systems is a fascinating subject which encompasses very diverse phenomena such as the fractional quantum Hall effect \cite{FQH}, high-temperature superconductivity \cite{HTSC1,HTSC2}, heavy-fermion compounds 
\cite{HF1,HF2}, etc. Despite a large effort of experimentalists and theoreticians, only recently has some progress on the quantitative understanding of these systems been achieved, in particular of the Mott transition in fermionic systems \cite{Mott}. The main difficulty lies in the fact that two vastly different energy scales play an important role in the redistribution of the spectral weight. A description of the low-energy physics 
requires very accurate understanding of the interplay with 
high-energy excitations dominated by short-range Coulomb repulsion. 
The interplay between low-energy spin fluctuations and the fermionic excitations is believed to ultimately lead to the phenomenon of high-temperature superconductivity \cite{HTSC2}. 
The development of reliable theoretical tools to calculate the material specific properties of strongly correlated materials remains one of the greatest challenges in modern theoretical condensed matter physics.

Dynamical mean-field theory (DMFT) \cite{DMFT}, which maps the strongly correlated lattice problem to a quantum impurity problem coupled to an electronic bath was a major step forward in the understanding of these systems. DMFT captures local temporal fluctuations, but does not incorporate spatial fluctuations. Its success is based on the fact that the physics is dominated by strong local interactions and that the phases of the fermions are averaged over a large number of bonds connecting a given site to the lattice, so that memory effects in the bath can be neglected, see e.g. \cite{Mueller-Hartmann:1984,MetznerVollhardt:1989}. 

Many efforts have been undertaken to go beyond the mean-field description. An expansion in $1/z$ where $z$ is the coordination number does not converge as the action depends in a non-analytic way on the coordination number \cite{schiller}. Using an auxiliary field it is possible to perform a systematic cumulant expansion of the path integral for the partition function around the strong coupling limit as pointed out in \cite{StanescuKotliar:2004}. There the authors developed a general framework for expanding around DMFT even considering non-local Coulomb interaction.
Cluster generalizations of DMFT \cite{cluster1,cluster2,cluster3,cluster4} break translational invariance either in real or momentum space explicitly and might artificially favor states which order at some finite wave-vector. 
The correlations included are necessarily short-ranged. Effects like  pseudogap formation and correlations related with a narrow region of {\it reciprocal} space, such as the vicinity of Van Hove singularities \cite{VH}, can hardly be taken into account by cluster approaches. 
A momentum dependent self-energy was introduced into DMFT by including nonlocal correlations through a classical fluctuating field \cite{pseudogap}.
Recently, straightforward  diagrammatic extensions to DMFT to include long-range correlations have been proposed \cite{toschi,Kus06}. 
It was further recognized, 
that a systematic expansion around DMFT can be transformed into a standard diagrammatic technique in terms of auxiliary, so-called dual fermions. 

In this Letter, we present results from a fully self-consistent infinite ladder diagram summation in terms of dual fermions. This approach manifestly takes long-wavelength correlations, in particular magnon contributions, into account.
We show that the convergence of the perturbation series is considerably enhanced by mapping the strongly correlated lattice fermions to weakly correlated dual fermions. This leads to a rapid convergence in the weak and strong coupling regimes and even far from these limits. Our study of the convergence properties is based on the leading eigenvalue analysis derived from the Bethe-Salpeter equation (BSE).

To be specific, we start with the one-band Hubbard model in two dimensions (2D)
\[
H=-t\sum_{ij,\sigma}c_{i\sigma}^{\dagger}c_{j\sigma}^{\phantom{\dagger}}+U\sum_{i}n_{i\uparrow}n_{i\downarrow}.
\]
Here $c_{i\sigma}^{\dagger}$ denotes the creation operator of an electron at site $i$ with spin $\sigma$, 
$t$ denotes the nearest neighbor hopping amplitude and $U$ is the strength of the on-site (screened) Coulomb interaction.

\begin{figure}[t]
\begin{centering}
\includegraphics[width=7.5cm]{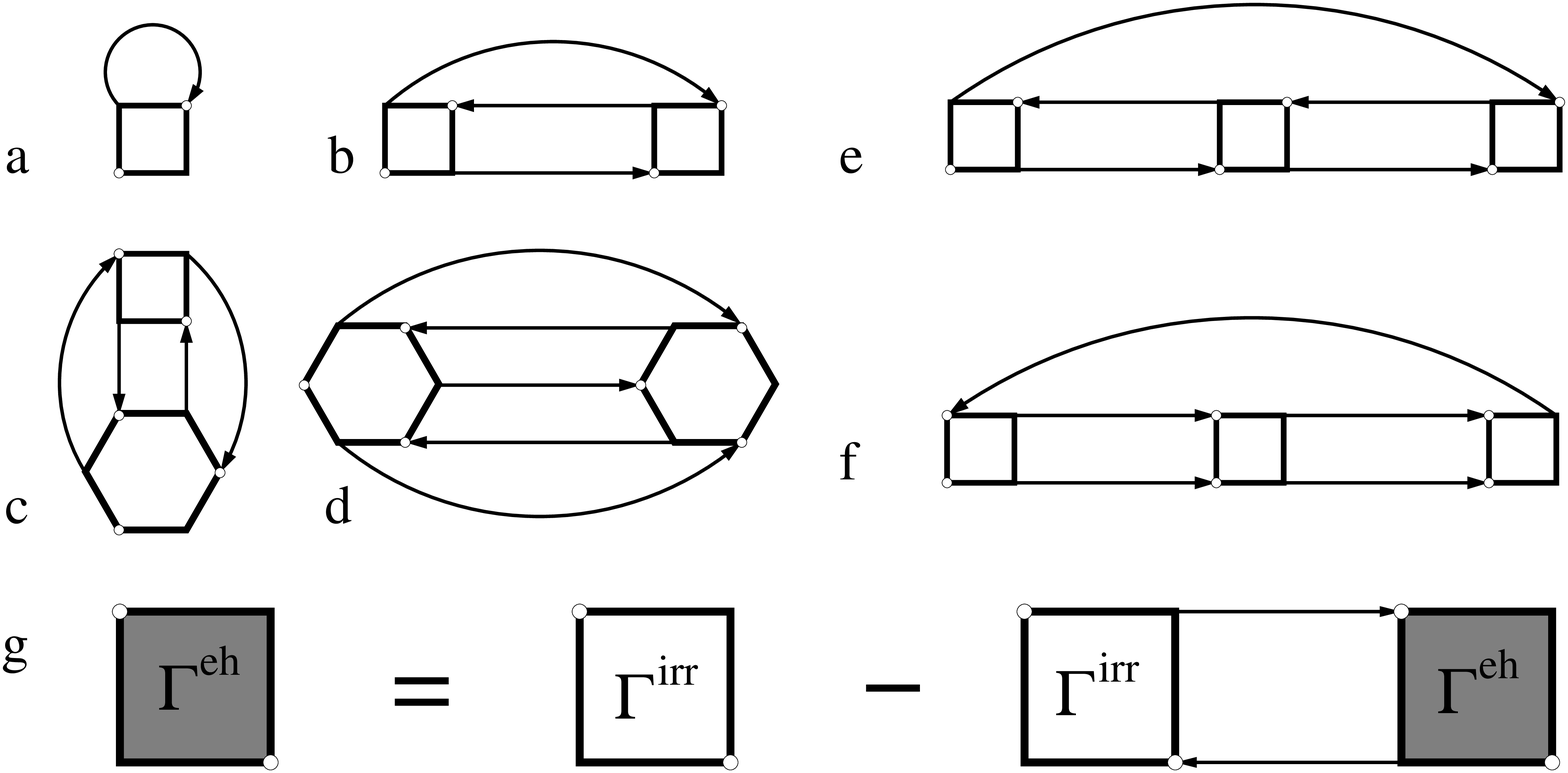}
\par\end{centering}
\caption{Diagrams for the dual self-energy $\Sigma^{d}$. The exact two- and three-particle impurity vertices are depicted by squares and hexagons. The lines are dual Green functions. Diagrams a) and c) are purely local, while the remaining diagrams give non-local contributions. g) Diagrammatic representation of the Bethe-Salpeter equation in the electron-hole channel.}
\label{fig::diagrams}
\end{figure}

The formulation of the theory for this model relies on a separation of the lattice action into two parts: an optimal impurity problem and a bilinear term that couples the impurities. This is achieved by introducing a dynamical field $\Delta_\omega$ at each lattice site. The resulting imaginary time action takes the form
\begin{equation*}
S[c^{\ast},c]=\sum_{i}S_{\text{imp}}[c_{\omega i\sigma}^{\ast},c_{\omega i\sigma}]+\sum_{\omega{\bf k}\sigma}c_{\omega{\bf  k}\sigma}^{\ast}\left(\epsilon_{{\bf k}}-\Delta_{\omega}\right)c_{\omega{\bf k}\sigma}\ .
\label{eqn::action}
\end{equation*}
Here $\epsilon_{\vc{k}}=-2t (\cos k_x + \cos k_y)$ is the bare dispersion with bandwidth $W=8t$, $\omega$ are the Matsubara frequencies and $\vc{k}$ labels momentum. $S_{\text{imp}}$ is the action for an impurity embedded in a time dependent electronic bath described by the hybridization function $\Delta_\omega$, which is the analogue of the dynamical field in DMFT. 
The effect of spatial correlations enters via the remainder of the lattice action and renders an exact solution impossible. This term is transformed by introducing auxiliary fermionic degrees of freedom (the dual fermions) in the path integral via a continuous Hubbard-Stratonovich transformation. Integrating out the lattice fermions produces the correlation functions of the impurity. These enter the dual potential and Green's functions and are obtained by  solving the impurity problem using a numerically exact continuous-time QMC algorithm \cite{Rub05-3}. 
The lattice problem is solved perturbatively in terms of the dual fermions through a standard diagrammatic expansion. The bare dual interactions correspond to the \emph{reducible} vertices of the impurity model,
which are connected by dual Green's functions as lines. In Fig. \ref{fig::diagrams} we show the lowest order diagrams. The rules to evaluate the diagrams are essentially those of the antisymmetrized diagrammatic technique \cite{AGD}.
More details on the method can be found in Refs. [\onlinecite{Rub08,brener,Rub09,Haf07,gangli}].

The ladder dual fermion approximation (LDFA) is obtained by functional derivative of a suitable dual Luttinger-Ward functional, $\Sigma_\text{LDFA}^\text{d} =\delta\Phi^\text{d}/\delta G^\text{d}$.
A one-to-one correspondence between the dual and original functionals \cite{Rub08} ensures that the theory is conserving in the Baym-Kadanoff sense \cite{BK}. 
It is important that \emph{irrespective} of the spatial dimension, DMFT appears as the lowest order approximation in this functional formulation.
We note that the LDFA takes two-particle excitations into account, which are the same for dual and lattice fermions \cite{brener,Rub09}.

The first three LDFA diagrams are shown in the first row of Fig. \ref{fig::diagrams} (a,b,e). We solve the BSE depicted in Fig. \ref{fig::diagrams} g) numerically. The LDFA self-energy is obtained from the Schwinger-Dyson equation, taking into account contributions from the horizontal and vertical particle-hole channels, similar to Ref. \cite{toschi}. Some care must be taken for combinatorial factors, to avoid overcounting of the second order contribution Fig. \ref{fig::diagrams} b), similar to the case encountered for Hugenholtz diagrams.

We assess the reliability of our method in the vicinity of the AFI, where long-range fluctuations are expected to be of vital importance.
Different approximations are compared by their leading eigenvalue of a linear eigenvalue problem derived from the BSE in the electron-hole channel of Fig. \ref{fig::diagrams} g).
For fixed transferred frequency $\Omega$ and momentum $\vc{q}$, it reads
\begin{equation}
-\frac{T}{N}\sum\limits
_{w'\vc{k}'}\Gamma_{\omega\omega'\Omega}^{\text{irr},s}\,
G_{\omega'}(\vc{k}')\,
G_{\omega'+\Omega}(\vc{k}'+\vc{q})\phi_{\omega'}=\lambda\phi_{\omega}\ .
\label{eqn::bse}
\end{equation}
Here $\Gamma^{s}=\Gamma^{\uparrow\uparrow}-\Gamma^{\uparrow\downarrow}$ denotes the singlet spin channel of the irreducible vertex. $T$ is the temperature and $N$ the number of k-points. We focus on the leading eigenvalues in the vicinity of the AFI and hence on $\vc{q}=(\pi,\pi)$ and $\Omega=0$. An eigenvalue of one indicates a  transition to the symmetry broken state.

\begin{figure}[t]
\begin{centering}
\includegraphics[width=7.5cm]{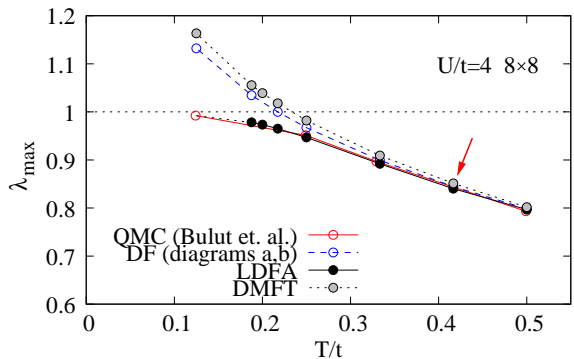}
\par\end{centering}
\caption{(Color online) Comparison of the leading eigenvalue of the Bethe-Salpeter equation in the electron-hole channel for a wavevector $\vc{q}=(\pi,\pi)$ and $\Omega=0$ obtained within different approximations with quantum Monte-Carlo data taken from Ref. \cite{bulut} for an $8\times 8$ lattice. An eigenvalue of $\lambda_\text{max}$ indicates the antiferromagnetic instability.}
\label{fig::qmc}
\end{figure}

For dual fermions, we have transformed the dual vertex $\Gamma^d_{\omega\omega'\Omega}(\vc{q})$ and Green's functions back to lattice quantities using an exact relation \cite{Rub08,brener}. The irreducible lattice vertex $\Gamma^{\text{irr}}_{\omega\omega'\Omega}(\vc{q})$ is obtained from the reducible one by inverting a BSE as depicted in Fig. \ref{fig::diagrams} g). Results are shown in Fig. \ref{fig::qmc}. While at higher temperatures all approximations give similar results, DMFT fails in the vicinity of the AFI. The instability at a N\'eel temperature of $T_{\text{N}}^{\text{DMFT}}/t=0.233$ is an artefact of the mean field approximation, which tends to stabilize the AF order.
Including short-range spatial correlations beyond DMFT, through the leading nonlocal diagram b) of Fig. \ref{fig::diagrams}, only slightly improves the solution and reduces the critical temperature down to $T_{\text{N}}^{\text{DF}}/t=0.215$.
This is in accordance to the leading eigenvalue being close to unity, indicating a decelerated convergence and pointing to the importance of long-wavelength fluctuations in the vicinity of the AFI.
This is facilitated through the LDFA, which complies with QMC even close to the AFI. This is remarkable because the results have been obtained perturbatively, starting from DMFT as a local approximation.

A self-consistent renormalization of the LDFA self-energy is essential below $T_{\text{N}}^{\text{DF}}$ as the eigenvalue of the BSE is larger than one in this regime and forbids a straightforward summation of the electron-hole ladder. Using a Green's function renormalized by the first few ladder diagrams pushes the eigenvalue below one and allows the self-consistent ladder summation. It is, however, not possible to approach the AFI arbitrarily close since it becomes computationally unfeasible to sum a large but finite number of ladder diagrams by iteratively solving the BSE. In the accessible temperature range, we do not find a sign of a transition to the ordered state.

In order to discuss the convergence properties, we consider the eigenvalue problem for dual fermions, by replacing Green's functions and the irreducible vertex by their dual counterparts (the impurity vertex for the latter) in Eq. \ref{eqn::bse}. Since the corresponding matrix is the building block of the electron-hole ladder, an eigenvalue equal to unity implies the divergence of the ladder sum and also a breakdown of the perturbation theory.

\begin{figure}[t]
\begin{centering}
\includegraphics[width=7.5cm]{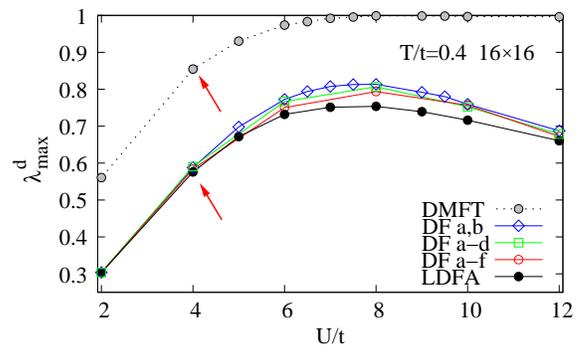}
\par\end{centering}
\caption{(Color online) Leading eigenvalue of the Bethe-Salpeter equation for dual fermions as function of $U$ in comparison to DMFT. The diagrams considered in each of the calculations are shown in the legend (cf. Fig. \ref{fig::diagrams}). 
The arrows point to the data for the same parameters as in Fig. \ref{fig::qmc}.}
\label{fig::lambdaU}
\end{figure}

The results are presented in Fig. \ref{fig::lambdaU}. For weak coupling, the leading eigenvalue is small and implies a fast convergence of the diagrams in the electron-hole ladder. 
More significantly, the eigenvalues decrease and converge to the same intercept in the large $U$ limit. This nicely illustrates that the dual perturbation theory smoothly interpolates between a standard perturbation expansion at small, and the cumulant expansion at large $U$, ensuring fast convergence in both regimes.
From the figure it is clear that this also improves the convergence properties for intermediate coupling ($U\sim W$). Even here corrections from approximations involving higher-order diagrams remain small, including those from the LDFA. Diagrams involving the three-particle vertex give a negligible contribution. Neglecting such diagrams is justified from a phase space argument \cite{AGD}. 

For a straightforward diagrammatic expansion around DMFT, the building block of the particle-hole ladder is constructed from the DMFT irreducible vertex and DMFT Green functions. As seen in Fig. \ref{fig::lambdaU}, the corresponding leading eigenvalue (and the effective interaction) is much larger than for dual fermions over the whole parameter range (e.g. at red arrows). This is also true for the leading eigenvalue of the lattice fermions (red arrow in Fig. \ref{fig::qmc}), which is close to the DMFT value for these parameters (the data labeled DMFT at the red arrows is the same in both plots). Remarkably, for the intermediate to strong coupling region, standard perturbation theory has to break down (since the eigenvalue approaches one), while for a theory in terms of dual fermions, this is not the case.

\begin{figure}[t]
\begin{centering}
\includegraphics[width=6.0cm]{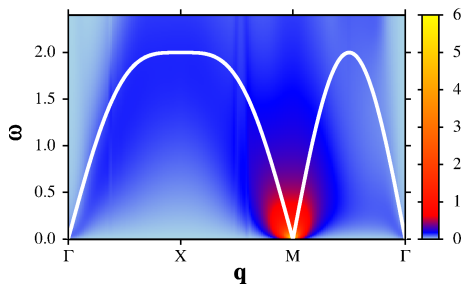}
\par\end{centering}
\caption{(Color online) Dynamical susceptibility $\chi(\vc{q},\omega)$ for $U/t=4$, and $T/t=0.19$.}
\label{fig::results}
\end{figure}

In Fig. \ref{fig::results} we show the dynamical susceptibility $\chi(\vc{q},\omega)$ obtained at half-filling for $U/t=4$ and $T/t=0.19$, together with the dispersion from spin-wave theory for a Heisenberg model with AF exchange $J=4t^2/U$.
It clearly displays the magnon spectrum in the paramagnetic state. We find a broadening of the frequency distribution and a small frequency shift of the peak from zero at the wave-vector $\vc{q}=(\pi,\pi)$ ($M$-point). Such a behavior is reminiscent of a 2D Heisenberg model at finite temperature where a short-range order with correlation length $\xi \gg a$ ($a$ is the lattice constant) takes place and a typical small energy scale of order $J a/\xi$ arises \cite{IK1}.

\begin{figure}[b]
\begin{centering}
\includegraphics[width=8.5cm]{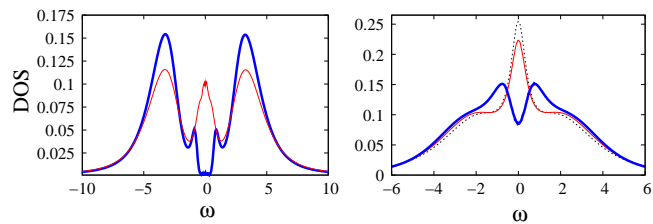}
\par\end{centering}
\caption{(Color online) Left: Metallic and insulating DOS within the coexistence region of the Mott transition $U/t=6.25$, $T/t=0.08$. Right: DMFT (dashed), DF (diagrams a,b) (thin line) and LDFA (thick line) DOS at $U/t=4$, $T/t=0.19$. The LDFA DOS exhibits the antiferromagnetic pseudogap.}
\label{fig::dos}
\end{figure}

Figure \ref{fig::dos} (left) shows the local density of states (DOS) at the Mott transition. The insulating (thick line) and metallic DOS (thin line) was calculated in the coexistence region for $U/t=6.25$ and $T/t=0.08$, where DMFT gives a metallic solution. The insulating solution exhibits the characteristic coherence peaks at the gap edge. The short-range AF correlations lead to a smeared ``antiferromagnetic gap''-like behavior \cite{IK1}. In our approach it is possible to account for the strong, temperature dependent modification of the DMFT hybridization function in a self-consistent procedure. This is essential since the Mott transition cannot be described perturbatively \cite{Mott}. We find a sizeable reduction of the critical $U$ at the second-order endpoint at $T/t\sim 0.11$ from $U_c=9.35$ in DMFT down to $U_c\sim 6.5$ due to the short-range AF correlations. The transition remains first order but displays a qualitative modification of the transition lines $U_{c1,2}(T)$ compared to DMFT, which is in accordance with recent dynamical cluster approximation results \cite{park}.


To further demonstrate the power of the method, we show the DOS for $U/t=4$ and $T/t=0.19$, obtained within DMFT (dashed line), dual fermion
approximation with diagram Fig. \ref{fig::diagrams} b) and the LDFA in Fig. \ref{fig::dos} (right) at half-filling. 
While the short-range correlations only slightly reduce the DOS at zero frequency, the pseudogap opens as long-range correlations are included through the LDFA. 
The pseudogap has been obtained within large scale cluster DMFT calculations \cite{cluster1}, in the symmetry broken state \cite{Rub09} or semiclassically \cite{pseudogap}. Here we obtain it for the first time within a translationally invariant solution in the paramagnetic state and for a $U$ well below the Mott transition point. A pseudogap behavior previously reported at $U/t=8$ \cite{Kus06} is reproduced using only diagram b) of Fig. \ref{fig::diagrams} and the DMFT hybridization, so that it is actually the precursor of the short-range AF correlation-assisted Mott transition.

To conclude, we have shown that the dual ladder approximation efficiently takes long-range correlations into account and is superior to straightforward diagrammatic expansions. The theory treats spatial correlations on all length scales on equal footing, is complementary to cluster approaches and can access low temperatures away from half filling. 
The multi-orbital formulation \cite{Haf07} in combination with density functional theory provides a computationally feasible scheme 
for a systematic study of strongly correlated real materials of present day interest, such as the high-$T_c$ materials, or quasi-one-dimensional organic conductors.

H.H. gratefully acknowledges valuable discussions with V. Jani\v{s} and M. Ringel. This work was supported by DFG Grant No. SFB 668-A3 and DFG project No. 436 RUS 113/938/0 (Germany), NWO project 047.016.005 and FOM (The Netherlands), the Leading
scientific schools program and RFFI grants 08-03-00930, 08-02-91953, 08-02-01020 (Russia).


\begin{thebibliography}{31}
\expandafter\ifx\csname natexlab\endcsname\relax\def\natexlab#1{#1}\fi
\expandafter\ifx\csname bibnamefont\endcsname\relax
  \def\bibnamefont#1{#1}\fi
\expandafter\ifx\csname bibfnamefont\endcsname\relax
  \def\bibfnamefont#1{#1}\fi
\expandafter\ifx\csname citenamefont\endcsname\relax
  \def\citenamefont#1{#1}\fi
\expandafter\ifx\csname url\endcsname\relax
  \def\url#1{\texttt{#1}}\fi
\expandafter\ifx\csname urlprefix\endcsname\relax\def\urlprefix{URL }\fi
\providecommand{\bibinfo}[2]{#2}
\providecommand{\eprint}[2][]{\url{#2}}

\bibitem[{\citenamefont{Prange and Girvin}(1997)}]{FQH}
\bibinfo{editor}{\bibfnamefont{R.~E.} \bibnamefont{Prange}} \bibnamefont{and}
  \bibinfo{editor}{\bibfnamefont{S.~M.} \bibnamefont{Girvin}}, eds.,
  \emph{\bibinfo{title}{The Quantum Hall Effect}}
  (\bibinfo{publisher}{Springer, New York}, \bibinfo{year}{1997}).

\bibitem[{\citenamefont{Anderson}(1997)}]{HTSC1}
\bibinfo{author}{\bibfnamefont{P.~W.} \bibnamefont{Anderson}},
  \emph{\bibinfo{title}{The Theory of Superconductivity in High-T$_c$
  Cuprates}} (\bibinfo{publisher}{Princeton Univ. Press, Princeton},
  \bibinfo{year}{1997}).

\bibitem[{\citenamefont{Scalapino}(1995)}]{HTSC2}
\bibinfo{author}{\bibfnamefont{D.~J.} \bibnamefont{Scalapino}},
  \bibinfo{journal}{Phys. Rep.} \textbf{\bibinfo{volume}{250}},
  \bibinfo{pages}{329} (\bibinfo{year}{1995}).

\bibitem[{\citenamefont{Stewart}(1984)}]{HF1}
\bibinfo{author}{\bibfnamefont{G.~R.} \bibnamefont{Stewart}},
  \bibinfo{journal}{Rev. Mod. Phys} \textbf{\bibinfo{volume}{56}},
  \bibinfo{pages}{755} (\bibinfo{year}{1984}).

\bibitem[{\citenamefont{v.~L\"{o}hneysen
  et~al.}(2007)\citenamefont{v.~L\"{o}hneysen, Rosch, Vojta, and
  W\"{o}lfle}}]{HF2}
\bibinfo{author}{\bibfnamefont{H.}~\bibnamefont{v.~L\"{o}hneysen}},
  \bibinfo{author}{\bibfnamefont{A.}~\bibnamefont{Rosch}},
  \bibinfo{author}{\bibfnamefont{M.}~\bibnamefont{Vojta}}, \bibnamefont{and}
  \bibinfo{author}{\bibfnamefont{P.}~\bibnamefont{W\"{o}lfle}},
  \bibinfo{journal}{Rev. Mod. Phys.} \textbf{\bibinfo{volume}{79}},
  \bibinfo{pages}{1015} (\bibinfo{year}{2007}).

\bibitem[{\citenamefont{Mott}(1974)}]{Mott}
\bibinfo{author}{\bibfnamefont{N.~F.} \bibnamefont{Mott}},
  \emph{\bibinfo{title}{Metal - Insulator Transitions}}
  (\bibinfo{publisher}{Taylor and Francis, London}, \bibinfo{year}{1974}).

\bibitem[{\citenamefont{Georges et~al.}(1996)\citenamefont{Georges, Kotliar,
  Krauth, and Rozenberg}}]{DMFT}
\bibinfo{author}{\bibfnamefont{A.}~\bibnamefont{Georges}},
  \bibinfo{author}{\bibfnamefont{G.}~\bibnamefont{Kotliar}},
  \bibinfo{author}{\bibfnamefont{W.}~\bibnamefont{Krauth}}, \bibnamefont{and}
  \bibinfo{author}{\bibfnamefont{M.}~\bibnamefont{Rozenberg}},
  \bibinfo{journal}{Rev. Mod. Phys.} \textbf{\bibinfo{volume}{68}},
  \bibinfo{pages}{13} (\bibinfo{year}{1996}).

\bibitem[{\citenamefont{M{\"u}ller-Hartmann}(1984)}]{Mueller-Hartmann:1984}
\bibinfo{author}{\bibfnamefont{E.}~\bibnamefont{M{\"u}ller-Hartmann}},
  \bibinfo{journal}{Z.~Phys.~B} \textbf{\bibinfo{volume}{57}},
  \bibinfo{pages}{281} (\bibinfo{year}{1984}).

\bibitem[{\citenamefont{Metzner and Vollhardt}(1989)}]{MetznerVollhardt:1989}
\bibinfo{author}{\bibfnamefont{W.}~\bibnamefont{Metzner}} \bibnamefont{and}
  \bibinfo{author}{\bibfnamefont{D.}~\bibnamefont{Vollhardt}},
  \bibinfo{journal}{Phys. Rev. Lett.} \textbf{\bibinfo{volume}{62}},
  \bibinfo{pages}{324} (\bibinfo{year}{1989}).

\bibitem[{\citenamefont{Schiller and Ingersent}(1995)}]{schiller}
\bibinfo{author}{\bibfnamefont{A.}~\bibnamefont{Schiller}} \bibnamefont{and}
  \bibinfo{author}{\bibfnamefont{K.}~\bibnamefont{Ingersent}},
  \bibinfo{journal}{Phys. Rev. Lett.} \textbf{\bibinfo{volume}{75}},
  \bibinfo{pages}{113} (\bibinfo{year}{1995}).

\bibitem[{\citenamefont{Stanescu and Kotliar}(2004)}]{StanescuKotliar:2004}
\bibinfo{author}{\bibfnamefont{T.~D.} \bibnamefont{Stanescu}} \bibnamefont{and}
  \bibinfo{author}{\bibfnamefont{G.}~\bibnamefont{Kotliar}},
  \bibinfo{journal}{Phys. Rev. B} \textbf{\bibinfo{volume}{70}},
  \bibinfo{pages}{205112} (\bibinfo{year}{2004}).

\bibitem[{\citenamefont{Maier et~al.}(2005)\citenamefont{Maier, Jarrell,
  Pruschke, and Hettler}}]{cluster1}
\bibinfo{author}{\bibfnamefont{T.}~\bibnamefont{Maier}},
  \bibinfo{author}{\bibfnamefont{M.}~\bibnamefont{Jarrell}},
  \bibinfo{author}{\bibfnamefont{T.}~\bibnamefont{Pruschke}}, \bibnamefont{and}
  \bibinfo{author}{\bibfnamefont{M.~H.} \bibnamefont{Hettler}},
  \bibinfo{journal}{Rev. Mod. Phys.} \textbf{\bibinfo{volume}{77}},
  \bibinfo{eid}{1027} (\bibinfo{year}{2005}).

\bibitem[{\citenamefont{Lichtenstein and Katsnelson}(2000)}]{cluster2}
\bibinfo{author}{\bibfnamefont{A.~I.} \bibnamefont{Lichtenstein}}
  \bibnamefont{and} \bibinfo{author}{\bibfnamefont{M.~I.}
  \bibnamefont{Katsnelson}}, \bibinfo{journal}{Phys. Rev. B}
  \textbf{\bibinfo{volume}{62}}, \bibinfo{pages}{R9283} (\bibinfo{year}{2000}).

\bibitem[{\citenamefont{Kotliar et~al.}(2001)\citenamefont{Kotliar, Savrasov,
  Palsson, and Biroli}}]{cluster3}
\bibinfo{author}{\bibfnamefont{G.}~\bibnamefont{Kotliar}},
  \bibinfo{author}{\bibfnamefont{S.~Y.} \bibnamefont{Savrasov}},
  \bibinfo{author}{\bibfnamefont{G.}~\bibnamefont{Palsson}}, \bibnamefont{and}
  \bibinfo{author}{\bibfnamefont{G.}~\bibnamefont{Biroli}},
  \bibinfo{journal}{Phys. Rev. Lett.} \textbf{\bibinfo{volume}{87}},
  \bibinfo{pages}{186401} (\bibinfo{year}{2001}).

\bibitem[{\citenamefont{Potthoff et~al.}(2003)\citenamefont{Potthoff, Aichhorn,
  and Dahnken}}]{cluster4}
\bibinfo{author}{\bibfnamefont{M.}~\bibnamefont{Potthoff}},
  \bibinfo{author}{\bibfnamefont{M.}~\bibnamefont{Aichhorn}}, \bibnamefont{and}
  \bibinfo{author}{\bibfnamefont{C.}~\bibnamefont{Dahnken}},
  \bibinfo{journal}{Phys. Rev. Lett.} \textbf{\bibinfo{volume}{91}},
  \bibinfo{pages}{206402} (\bibinfo{year}{2003}).

\bibitem[{\citenamefont{Irkhin et~al.}(2002)\citenamefont{Irkhin, Katanin, and
  Katsnelson}}]{VH}
\bibinfo{author}{\bibfnamefont{V.~Y.} \bibnamefont{Irkhin}},
  \bibinfo{author}{\bibfnamefont{A.~A.} \bibnamefont{Katanin}},
  \bibnamefont{and} \bibinfo{author}{\bibfnamefont{M.~I.}
  \bibnamefont{Katsnelson}}, \bibinfo{journal}{Phys. Rev. Lett.}
  \textbf{\bibinfo{volume}{89}}, \bibinfo{pages}{076401}
  (\bibinfo{year}{2002}).

\bibitem[{\citenamefont{Sadovskii et~al.}(2005)\citenamefont{Sadovskii,
  Nekrasov, Kuchinskii, Pruschke, and Anisimov}}]{pseudogap}
\bibinfo{author}{\bibfnamefont{M.~V.} \bibnamefont{Sadovskii}},
  \bibinfo{author}{\bibfnamefont{I.~A.} \bibnamefont{Nekrasov}},
  \bibinfo{author}{\bibfnamefont{E.~Z.} \bibnamefont{Kuchinskii}},
  \bibinfo{author}{\bibfnamefont{T.}~\bibnamefont{Pruschke}}, \bibnamefont{and}
  \bibinfo{author}{\bibfnamefont{V.~I.} \bibnamefont{Anisimov}},
  \bibinfo{journal}{Phys. Rev. B} \textbf{\bibinfo{volume}{72}},
  \bibinfo{pages}{155105} (\bibinfo{year}{2005}).

\bibitem[{\citenamefont{Toschi et~al.}(2007)\citenamefont{Toschi, Katanin, and
  Held}}]{toschi}
\bibinfo{author}{\bibfnamefont{A.}~\bibnamefont{Toschi}},
  \bibinfo{author}{\bibfnamefont{A.~A.} \bibnamefont{Katanin}},
  \bibnamefont{and} \bibinfo{author}{\bibfnamefont{K.}~\bibnamefont{Held}},
  \bibinfo{journal}{Phys. Rev. B} \textbf{\bibinfo{volume}{75}},
  \bibinfo{eid}{045118} (\bibinfo{year}{2007}).

\bibitem[{\citenamefont{Kusunose}(2006)}]{Kus06}
\bibinfo{author}{\bibfnamefont{H.}~\bibnamefont{Kusunose}},
  \bibinfo{journal}{J. Phys. Soc. Jpn.} \textbf{\bibinfo{volume}{75}},
  \bibinfo{pages}{054713} (\bibinfo{year}{2006}).

\bibitem[{\citenamefont{Rubtsov et~al.}(2005)\citenamefont{Rubtsov, Savkin, and
  Lichtenstein}}]{Rub05-3}
\bibinfo{author}{\bibfnamefont{A.~N.} \bibnamefont{Rubtsov}},
  \bibinfo{author}{\bibfnamefont{V.~V.} \bibnamefont{Savkin}},
  \bibnamefont{and} \bibinfo{author}{\bibfnamefont{A.~I.}
  \bibnamefont{Lichtenstein}}, \bibinfo{journal}{Phys. Rev. B}
  \textbf{\bibinfo{volume}{72}}, \bibinfo{pages}{035122}
  (\bibinfo{year}{2005}).

\bibitem[{\citenamefont{Rubtsov et~al.}(2008)\citenamefont{Rubtsov, Katsnelson,
  and Lichtenstein}}]{Rub08}
\bibinfo{author}{\bibfnamefont{A.~N.} \bibnamefont{Rubtsov}},
  \bibinfo{author}{\bibfnamefont{M.~I.} \bibnamefont{Katsnelson}},
  \bibnamefont{and} \bibinfo{author}{\bibfnamefont{A.~I.}
  \bibnamefont{Lichtenstein}}, \bibinfo{journal}{Phys. Rev. B}
  \textbf{\bibinfo{volume}{77}}, \bibinfo{pages}{033101}
  (\bibinfo{year}{2008}).

\bibitem[{\citenamefont{Brener et~al.}(2008)\citenamefont{Brener, Hafermann,
  Rubtsov, Katsnelson, and Lichtenstein}}]{brener}
\bibinfo{author}{\bibfnamefont{S.}~\bibnamefont{Brener}},
  \bibinfo{author}{\bibfnamefont{H.}~\bibnamefont{Hafermann}},
  \bibinfo{author}{\bibfnamefont{A.~N.} \bibnamefont{Rubtsov}},
  \bibinfo{author}{\bibfnamefont{M.~I.} \bibnamefont{Katsnelson}},
  \bibnamefont{and} \bibinfo{author}{\bibfnamefont{A.~I.}
  \bibnamefont{Lichtenstein}}, \bibinfo{journal}{Phys. Rev. B}
  \textbf{\bibinfo{volume}{77}}, \bibinfo{eid}{195105} (\bibinfo{year}{2008}).

\bibitem[{\citenamefont{Rubtsov et~al.}(2009)\citenamefont{Rubtsov, Katsnelson,
  Lichtenstein, and Georges}}]{Rub09}
\bibinfo{author}{\bibfnamefont{A.~N.} \bibnamefont{Rubtsov}},
  \bibinfo{author}{\bibfnamefont{M.~I.} \bibnamefont{Katsnelson}},
  \bibinfo{author}{\bibfnamefont{A.~I.} \bibnamefont{Lichtenstein}},
  \bibnamefont{and} \bibinfo{author}{\bibfnamefont{A.}~\bibnamefont{Georges}},
  \bibinfo{journal}{Phys. Rev. B} \textbf{\bibinfo{volume}{79}},
  \bibinfo{eid}{045133} (\bibinfo{year}{2009}).

\bibitem[{\citenamefont{Hafermann et~al.}(2007)\citenamefont{Hafermann, Brener,
  Rubtsov, Katsnelson, and Lichtenstein}}]{Haf07}
\bibinfo{author}{\bibfnamefont{H.}~\bibnamefont{Hafermann}},
  \bibinfo{author}{\bibfnamefont{S.}~\bibnamefont{Brener}},
  \bibinfo{author}{\bibfnamefont{A.~N.} \bibnamefont{Rubtsov}},
  \bibinfo{author}{\bibfnamefont{M.~I.} \bibnamefont{Katsnelson}},
  \bibnamefont{and} \bibinfo{author}{\bibfnamefont{A.~I.}
  \bibnamefont{Lichtenstein}}, \bibinfo{journal}{JETP Lett.}
  \textbf{\bibinfo{volume}{86}}, \bibinfo{pages}{677} (\bibinfo{year}{2007}).

\bibitem[{\citenamefont{Li et~al.}(2008)\citenamefont{Li, Lee, and
  Monien}}]{gangli}
\bibinfo{author}{\bibfnamefont{G.}~\bibnamefont{Li}},
  \bibinfo{author}{\bibfnamefont{H.}~\bibnamefont{Lee}}, \bibnamefont{and}
  \bibinfo{author}{\bibfnamefont{H.}~\bibnamefont{Monien}},
  \bibinfo{journal}{Phys. Rev. B} \textbf{\bibinfo{volume}{78}},
  \bibinfo{eid}{195105} (\bibinfo{year}{2008}).

\bibitem[{\citenamefont{Abrikosov et~al.}(1965)\citenamefont{Abrikosov,
  Gor'kov, and Dzyaloshinskii}}]{AGD}
\bibinfo{author}{\bibfnamefont{A.~A.} \bibnamefont{Abrikosov}},
  \bibinfo{author}{\bibfnamefont{L.~P.} \bibnamefont{Gor'kov}},
  \bibnamefont{and} \bibinfo{author}{\bibfnamefont{I.~E.}
  \bibnamefont{Dzyaloshinskii}}, \emph{\bibinfo{title}{Methods of Quantum Field
  Theory in Statistical Physics}} (\bibinfo{publisher}{Pergamon Press, New
  York}, \bibinfo{year}{1965}).

\bibitem[{\citenamefont{Baym and Kadanoff}(1961)}]{BK}
\bibinfo{author}{\bibfnamefont{G.}~\bibnamefont{Baym}} \bibnamefont{and}
  \bibinfo{author}{\bibfnamefont{L.~P.} \bibnamefont{Kadanoff}},
  \bibinfo{journal}{Phys. Rev.} \textbf{\bibinfo{volume}{124}},
  \bibinfo{pages}{287} (\bibinfo{year}{1961}).

\bibitem[{\citenamefont{Bulut et~al.}(1993)\citenamefont{Bulut, Scalapino, and
  White}}]{bulut}
\bibinfo{author}{\bibfnamefont{N.}~\bibnamefont{Bulut}},
  \bibinfo{author}{\bibfnamefont{D.~J.} \bibnamefont{Scalapino}},
  \bibnamefont{and} \bibinfo{author}{\bibfnamefont{S.~R.} \bibnamefont{White}},
  \bibinfo{journal}{Phys. Rev. B} \textbf{\bibinfo{volume}{47}},
  \bibinfo{pages}{14599} (\bibinfo{year}{1993}).

\bibitem[{\citenamefont{Irkhin and Katsnelson}(1991)}]{IK1}
\bibinfo{author}{\bibfnamefont{V.~Y.} \bibnamefont{Irkhin}} \bibnamefont{and}
  \bibinfo{author}{\bibfnamefont{M.~I.} \bibnamefont{Katsnelson}},
  \bibinfo{journal}{J. Phys. Condens. Matter} \textbf{\bibinfo{volume}{3}},
  \bibinfo{pages}{6439} (\bibinfo{year}{1991}).

\bibitem[{\citenamefont{Park et~al.}(2008)\citenamefont{Park, Haule, and
  Kotliar}}]{park}
\bibinfo{author}{\bibfnamefont{H.}~\bibnamefont{Park}},
  \bibinfo{author}{\bibfnamefont{K.}~\bibnamefont{Haule}}, \bibnamefont{and}
  \bibinfo{author}{\bibfnamefont{G.}~\bibnamefont{Kotliar}},
  \bibinfo{journal}{Phys. Rev. Lett.} \textbf{\bibinfo{volume}{101}},
  \bibinfo{eid}{186403} (\bibinfo{year}{2008}).

\end{thebibliography}
\end{document}